\documentclass[
    ,final            
  ]
  {aipproc}

\layoutstyle{6x9}
\IfFileExists{srcltx.sty}{\usepackage[active]{srcltx}}%

\begin{document}

\title{Sterile neutrinos}

\classification{14.60.St,13.15.+g,14.60.Pq,95.35.+d \hfill UCLA/07/TEP/5} 
\keywords      {}

\author{Alexander Kusenko}{
  address={Department of Physics and Astronomy, University of California, Los
Angeles, CA 90095-1547}
}

\begin{abstract}
Neutrino masses are usually described by adding to the Standard Model
some SU(2)-singlet fermions that have the Yukawa couplings, as well as some
Majorana mass terms.  The number of such fields and the scales of their
Majorana masses are not known.  Several independent observations point to the
possibility that some of these singlets may have masses well below the
electroweak scale.  A sterile neutrino with mass of a few keV can account
for cosmological dark matter.  The same particle would be emitted
anisotropically from a cooling neutron star born in a supernova explosion. 
This anisotropy can be large enough to explain the observed velocities of
pulsars. A lighter sterile neutrino, with mass of the order of eV, is implied
by the LSND results; it can have profound implications for cosmology.  We
review the physics of sterile neutrinos and the roles they may play in
astrophysics and cosmology.

\end{abstract}

\maketitle


\section{Sterile neutrinos in particle physics}

The name {\em sterile neutrino} was coined by Bruno~Pontecorvo, who
hypothesized the existence of the right-handed neutrinos in a seminal
paper~\cite{pontecorvo}, in which he also considered vacuum neutrino
oscillations in the laboratory and in astrophysics, the lepton number
violation, the neutrinoless double beta decay, some rare processes, such as
$\mu \rightarrow e \gamma$, and several other questions that have dominated the
neutrino physics for the next four decades.   Most models of the neutrino
masses introduce sterile (or right-handed) neutrinos to generate the masses of
the ordinary neutrinos via the seesaw mechanism~\cite{seesaw}.  The seesaw
lagrangian 
\begin{equation}
{\cal L}
  = {\cal L_{\rm SM}}+\bar N_{a} \left(i \gamma^\mu \partial_\mu 
\right )
N_{a}
  - y_{\alpha a} H \,  \bar L_\alpha N_{a} 
  - \frac{M_{a}}{2} \; \bar {N}_{a}^c N_{a} + h.c. \,,
\label{lagrangianM}
\end{equation}  
where ${\cal L_{\rm SM}}$ is the lagrangian of the Standard
Model, includes some number $n$  of singlet neutrinos $N_a$ ($a=1,...,n$) 
with Yukawa couplings $ y_{\alpha
a}$.  Here $H$ is the Higgs doublet and $L_\alpha$
($\alpha=e,\mu,\tau$) are the lepton doublets. Theoretical considerations do
not constrain the number $n$ of sterile neutrinos. In particular, there is no
constraint based on the anomaly cancellation because the sterile fermions do
not couple to the gauge fields. The experimental limits exist only for the
larger mixing angles~\cite{sterile_constraints}. To explain the neutrino masses
inferred from the atmospheric and solar neutrino experiments, $n=2$ singlets
are sufficient~\cite{2right-handed}, but a greater number is required if the
lagrangian (\ref{lagrangianM}) is to explain the LSND~\cite{deGouvea:2005er},
the r-process nucleosynthesis~\cite{r}, the pulsar 
kicks~\cite{ks97,fkmp,Kusenko:review}, dark
matter~\cite{dw,Fuller,shi_fuller,nuMSM}, and the formation of supermassive
black holes~\cite{biermann_munyaneza}.  

The scale of the right-handed Majorana masses $M_{a}$ is unknown; it can be
much greater than the electroweak scale~\cite{seesaw}, or it may be as low as a
few eV~\cite{deGouvea:2005er,nuMSM,deGouvea:2006gz}. Even if some of the
right-handed Majorana masses are much larger than others, for example, if some
of the $M_a$ ($a=1,...,n_{l}$) are smaller than 100~GeV, while some others
($a=n_{l},...,n$) are much greater than 100~GeV, both classes can have a
non-negligible contribution to the active neutrino masses. Obviously, this does
not contradict the usual decoupling theorems, because the heavy states decouple
from all the physical processes at low energies, but they can still contribute
to the values of the active neutrino masses if the corresponding Yukawa
couplings are large enough.

\subsection{Are they natural?} 

The seesaw mechanism~\cite{seesaw} can explain the smallness of the neutrino
masses in the presence of the Yukawa couplings of order one if the
Majorana masses $M_a$ are much larger than the electroweak scale. Indeed, in
this case the masses of the lightest neutrinos are suppressed by the ratios $
\langle H \rangle/M_a$.  

However, the origin of the Yukawa couplings remains unknown, and there is no
experimental evidence to suggest that these couplings must be of order 1. In
fact, the Yukawa couplings of the charged leptons are much smaller than 1. For
example, the Yukawa coupling of the electron is as small as $10^{-6}$.  
One can ask whether some theoretical models are more likely to produce the
numbers of order one or much smaller than one.  The two possibilities are, in
fact, realized in two types of theoretical models.  If the Yukawa couplings
arise as some topological intersection numbers in string theory, they are
generally expected to be of order one~\cite{Candelas:1987rx}, although very
small couplings are also possible~\cite{Eyton-Williams:2005bg}. If the Yukawa
couplings arise from the overlap of the wavefunctions of fermions located on
different branes in extra dimensions, they can be exponentially suppressed and
are expected to be very small~\cite{Mirabelli:1999ks}.  

In the absence of the fundamental theory, one may hope to gain some insight 
about the size of the Yukawa couplings using 't~Hooft's naturalness
criterion~\cite{tHooft}, which states essentially that a number can be
naturally small if setting it to zero increases the symmetry of the lagrangian.
 A small breaking of the symmetry is then associated with the small non-zero
value of the parameter.  This naturalness criterion has been applied to a
variety of theories; it is, for example, one of the main arguments in favor of
supersymmetry. (Setting the Higgs mass to zero does not increase the symmetry
of the Standard Model.  Supersymmetry relates the Higgs mass to the Higgsino
mass, which is protected by the chiral symmetry.  Therefore, the light Higgs
boson, which is not natural in the Standard Model, becomes natural in theories
with softly broken supersymmetry.) In view of 't~Hooft's criterion, the
\textit{small} Majorana mass is natural because setting $M_a$ to zero increases
the symmetry of the lagrangian
(\ref{lagrangianM})~\cite{Fujikawa:2004jy,deGouvea:2005er}.  

One can ask whether cosmology can provide any clues as to whether the mass
scale of sterile neutrinos should be above or below the electroweak scale.  It
is desirable to have a theory that could generate the matter--antimatter
asymmetry of the universe. In both limits of large and small $M_a$ one can have
a successful leptogenesis: in the case of the high-scale seesaw, the baryon
asymmetry can be generated from the out-of-equilibrium decays of heavy
neutrinos~\cite{Fukugita:1986hr}, while in the case of the low-energy seesaw,
the matter-antimatter asymmetry can be produced by  
the neutrino oscillations~\cite{baryogenesis}.  The
Big-Bang nucleosynthesis (BBN) can provide a constraint on the number of light
relativistic species in equilibrium~\cite{bbn,BBN}, but the sterile neutrinos
with the small mixing angles may never be in equilibrium in the early universe,
even at the highest temperatures~\cite{dw}.  Indeed, the effective mixing angle
of neutrinos at high temperature is suppressed due to the interactions with
plasma~\cite{high-T}, and, therefore, the sterile neutrinos may never
thermalize.  High-precision measurements of the primordial abundances may probe
the existence of sterile neutrinos and the lepton asymmetry of the universe in
the future~\cite{Smith:2006uw}.  

While many seesaw models assume that the sterile neutrinos have very large
masses, which makes them unobservable, it is worthwhile to consider light
sterile neutrinos in view of the above arguments, and also because they can
explain several experimental results.  In particular, sterile neutrinos can
account for cosmological dark matter~\cite{dw}, they can explain the observed
velocities of pulsars~\cite{ks97,fkmp,Kusenko:review}, the x-ray photons from
their decays can affect the star formation~\cite{reion}.  Finally, sterile
neutrinos can explain the LSND
result~\cite{deGouvea:2005er,Sorel:2003hf,LSND_sterile_decay},
which is currently being tested by the MiniBooNE experiment.

\section{Experimental status}

Laboratory experiments are able to set limits or discover sterile neutrinos
with a large enough mixing angle.  Depending on the mass, they
can be searched in different experiments. 

The light sterile neutrinos, with masses below $10^2$~eV, can be discovered in
one of the neutrino oscillations experiments~\cite{Smirnov:2006bu}.  In fact,
LSND has reported a result~\cite{LSND}, which, in combination with the other
experiments, implies the existence of at least one sterile neutrino, more
likely, two sterile
neutrinos~\cite{deGouvea:2005er,Sorel:2003hf}.  It is
also possible that sterile neutrino decays, rather than oscillations, are the 
explanation of the LSND result~\cite{LSND_sterile_decay}.

In the eV to MeV mass range, the ``kinks'' in the spectra of beta-decay
electrons can be used to set limits on sterile neutrinos mixed with the
electron neutrinos~\cite{Shrock81}.  Neutrinoless double beta decays can probe
the Majorana neutrino masses~\cite{Elliott:2002xe}.  An interesting
proposal is to search for sterile neutrinos in beta decays using a complete
kinematic reconstruction of the final state~\cite{Finocchiaro:1992hy}. 

For masses in the MeV--GeV range, peak searches in production of neutrinos
provide the limits.  The massive neutrinos $\nu_i$, if they exist, can be 
produced in meson decays, e.g. $\pi^\pm \rightarrow \mu^\pm \nu_i$,  with
probabilities that depend on the mixing in the charged current.  The
energy spectrum of muons in such decays should contain monochromatic
lines~\cite{Shrock81} at
$ 
T_i = ( m_\pi^2 + m_\mu^2 - 2 m_\pi m_\mu - m_{\nu_i}^2) / 2 m_\pi. 
$ 
Also, for the MeV--GeV masses one can set a number of constraints based on the
decays of the heavy neutrinos into the ``visible'' particles, which would
be observable by various detectors. These limits are discussed in
Ref.~\cite{sterile_constraints}.

\section{Sterile neutrinos in astrophysics and cosmology}

Sterile neutrinos can be produced in the early universe, as well as in
supernova explosions.  The light sterile neutrino, consistent with the LSND
result, is consistent with the existing bounds on the big-bang 
nucleosynthesis~\cite{bbn,BBN} and large-scale structure, especially if the
mixing lepton asymmetry of the universe is larger than the baryon
asymmetry~\cite{Chu:2006ua}. A heavier sterile neutrino, with mass in the keV
range is an appealing dark-matter candidate, as discussed below.

\begin{figure}
  \includegraphics[height=.5\textheight]{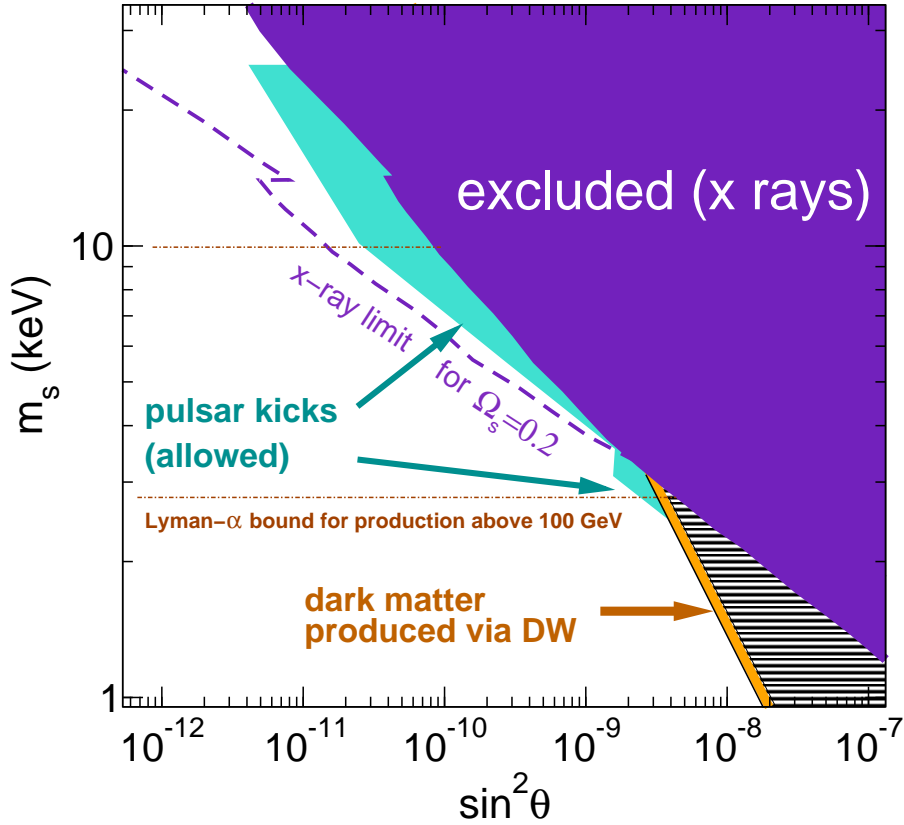}
  \caption{Sterile neutrinos with masses 2--25 keV can explain the pulsar
kicks if the mixing angles are large enough (as shown).  In the
region marked \textit{excluded (x-rays)}, the relic sterile neutrinos produced
in neutrino oscillations via the Dodelson--Widrow (DW) mechanism would have a
density inconsistent with the existing x-ray bounds.  If the sterile neutrinos
constitute all the dark matter, their masses and mixings should fall below the
dashed line. Note that DW mechanism is not sufficient to produce enough dark
matter for points on the dashed line: a large lepton
asymmetry~\cite{shi_fuller} or a new production
mechanism~\cite{shaposhnikov_tkachev,Kusenko:2006rh} is required. The
Lyman-$\alpha$ bound for dark-matter sterile neutrinos produced at temperatures
$T>100$~GeV is $m_s>2.7$~keV (see text or Ref.~\cite{Kusenko:2006rh} for
discussion).  The cosmological and the x-ray bounds do not apply if the
universe was never reheated above $T\sim$~MeV~\cite{low-reheat}. 
}
\label{fig:range}
\end{figure}

\subsection{Dark matter in the form of sterile neutrinos}

The sterile neutrinos can be the cosmological dark
matter~\cite{dw,Fuller,shi_fuller,nuMSM}.  The interactions already
present in the lagrangian (\ref{lagrangianM}) allow for the production of relic
sterile neutrinos via the Dodelson-Widrow (DW) mechanism~\cite{dw} in the right
amount to account for all dark matter, i.e. $\Omega_s\approx 0.2$, if one of
the Majorana masses is of the order of a keV.  We will denote the dark-matter
sterile neutrino $\nu_s$ and its mass $m_s$.  By assumption, the mixing angles
are small, and 
\begin{eqnarray}
  m_s & \approx & M_1  \\
\sin \theta & \approx &  \frac{y \langle H \rangle}{M_1}.
\end{eqnarray}

The mass and mixing angle are subject to the x-ray limits on the photons from
the decays of the relic sterile neutrinos~\cite{x-rays}, as well as the
Lyman-$\alpha$ bound~\cite{viel} discussed below (see
Fig.~\ref{fig:range}).

As was mentioned above, the relic sterile neutrinos can decay into the lighter
neutrinos and an the x-ray photons~\cite{pal_wolf}, which can be detected by
the x-ray telescopes~\cite{x-rays}. 
The flux of x-rays depends on the sterile neutrino abundance.  If all the dark
matter is made up of sterile neutrinos, $ \Omega_s\approx 0.2 $, then the
limit on the mass and the mixing angle is given by the dashed line in
Fig.~\ref{fig:range}. 
However, the interactions in the lagrangian (\ref{lagrangianM}) cannot produce
such an $ \Omega_s= 0.2 $ population of the sterile neutrinos for the masses
and mixing angles along this dashed line, unless the universe has a relatively
large lepton asymmetry~\cite{shi_fuller}.  If the lepton asymmetry is small,
the interactions in eq.~(\ref{lagrangianM}) can produce the relic sterile
neutrinos via the neutrino oscillations off-resonance at some sub-GeV
temperature~\cite{dw}. This mechanism provides the lowest possible abundance
(except for the low-temperature cosmologies, in which the universe is never
reheated above a few MeV after inflation~\cite{low-reheat}).  The
model-independent bound~\cite{Kusenko:2006rh} based on this scenario is shown
as a solid (purple) region in Fig.~\ref{fig:range}.  It is based on the flux
limit from Ref.~\cite{x-rays} and the analytical fit to the numerical
calculation of the sterile neutrino production by
Abazajian~\cite{Abazajian:2005gj}.  This calculation may have some
hadronic uncertainties~\cite{Asaka:2006rw}, but they appear to be under
control for the mass and mixing angle in the range of
interest~\cite{Asaka:2006nq}. 

If the lepton asymmetry of the universe is relatively large, the resonant
oscillations can produce the requisite amount of dark matter even for smaller
mixing angles~\cite{shi_fuller}, for which the x-ray limits are weak. (The
x-ray flux is proportional to the square of the mixing angle.) It is also
possible that some additional interactions, not present in
eq.~(\ref{lagrangianM}) can be responsible for the production of
dark-matter sterile neutrinos~\cite{shaposhnikov_tkachev,Kusenko:2006rh}.  We
will discuss this possibility in more detail below.

 The x-ray photons from sterile neutrino decays in the early universe could
have affected the star formation.  Although these x-rays alone are not
sufficient to reionize the universe, they can catalyze
the production of molecular hydrogen and speed up the star
formation~\cite{reion}, which, in turn, could cause the reionization.
Molecular hydrogen is a very important
cooling agent necessary for the collapse of primordial gas clouds that gave
birth to the first stars.  The fraction of molecular hydrogen must exceed a
certain minimal value for the star formation to begin~\cite{Tegmark:1996yt}. 
The reaction H+H$\rightarrow$H$_2 +\gamma$ is very slow in comparison with the
combination of reactions 
\begin{eqnarray}
{\rm H}^{+}+{\rm H}  & \rightarrow & {\rm H}_2^++ \gamma , \\  
{\rm H}_2^{+}+{\rm H} & \rightarrow & {\rm H}_2+{\rm H}^+, 
\end{eqnarray} %
which are possible if the hydrogen is ionized.  Therefore, the ionization
fraction determines the rate of molecular hydrogen production.  If dark
matter is made up of sterile neutrinos, their decays produce a sufficient flux
of photons to increase the ionization fraction by as much as two orders of
magnitude~\cite{reion}.  This has a dramatic effect on the
production of molecular hydrogen and the subsequent star formation.

Decays of the relic sterile neutrinos during the dark ages could produce an
observable signature in the 21-cm background~\cite{Valdes:2007cu}. It can be
detected and studied by such instruments as the Low Frequency Array (LOFAR),
the 21 Centimeter Array (21CMA), the Mileura Wide-field Array (MWA) and the
Square Kilometer Array~(SKA). 

\subsubsection{New physics at the electroweak scale, and the Lyman-$\alpha$ 
bounds}

One can ask whether the mass $M\sim$~keV in equation (\ref{lagrangianM}) is 
a fundamental constant of nature, or whether it could arise from 
some symmetry breaking via the Higgs mechanism.  For example, let us consider
the following modification of the lagrangian (\ref{lagrangianM}) following
Ref.~\cite{Kusenko:2006rh}: 
\begin{equation} 
{\cal L}
   =   {\cal L}_{0}+\bar N_{a} \left(i \gamma^\mu \partial_\mu 
\right )
N_{a}  - y_{\alpha a} H \,  \bar L_\alpha N_{a}  - \frac{h_a}{2} \, S \,
\bar {N}_{a}^c N_{a} 
 +V(H,S) + h.c. \,, 
\label{lagrangianS}
\end{equation}
where $ {\cal L}_{0}$ includes the gauge and kinetic terms of the Standard
Model, $H$ is the Higgs doublet, $S$ is the real boson, which is SU(2)-singlet,
$L_\alpha$ ($\alpha=e,\mu,\tau$) are the lepton doublets, and $ N_{a}$
($a=1,...,n$) are the additional singlet neutrinos.  Let us consider the
following scalar potential: 
\begin{equation}
V(H,S) =   m_{1}^2 |H|^2 + m_{2}^2 S^2+ \lambda_3 S^3 +  
\lambda_{_{HS}} |H|^2 S^2+ \lambda_{_S}  S^4   + \lambda_{_H}
|H|^4 .
\label{potential}
\end{equation}

After the symmetry breaking, the Higgs doublet and singlet fields each develop
a VEV, $\langle H\rangle= v_0=247$~GeV, $\langle S\rangle= v_1$, and the
singlet neutrinos acquire the Majorana masses $ M_a = h_a v_1$.  The mass of
the $S$ boson in after the symmetry breaking is $\tilde{m}_{_S}\sim v_1$.   The
presence of the singlet in the Higgs sector can be tested at the
LHC~\cite{LHC}.

This modification makes no difference in the low-energy theory, for example,
in its application to the masses of active neutrinos.  However, the new
coupling opens a new
channel for production of sterile neutrinos in the early universe.  Indeed,
if the couplings of $S$ to $H$ are large enough, while $h<10^{-6}$, the $S$
boson can be in equilibrium at temperatures above its mass, while the sterile
neutrino with a small mixing angle can be out of equilibrium at all times. 
This is the case, as long as the annihilations $NN \rightarrow NN$, $NN
\rightarrow {\rm scalars}$, etc.  are not fast enough to keep the sterile
neutrinos in equilibrium.  Now, since $S$ is in thermal equilibrium at
high temperatures, some amount of sterile neutrinos can be produced through
decays $S\rightarrow NN$.  The amount of sterile neutrinos produced this way is
determined by the $h$ coupling (and is independent of the active-sterile
neutrino mixing angle): 
\begin{equation}
\Omega_s =  0.2 \left ( \frac{33}{\xi} \right )
\left ( \frac{h}{ 1.4 \times 10^{-8} } \right )^3
\left ( \frac{ \langle S \rangle }{\tilde{m}_{_S} } \right )
\label{Omega_w_VEVs}, 
\end{equation}
where $\xi $ is the change in the number density of sterile neutrinos
relative to $T^3$ due to the dilution taking place as the universe cools. For
example, in the Standard Model, the reduction in the number of effective
degrees of freedom that occurs during the cooling from the temperature $T\sim
100$~GeV to a temperature below 1~MeV causes the entropy increase and the
dilution of any species out of equilibrium by factor $\xi\approx 33$. 

At the same time, the sterile neutrino mass is determined by the VEV of $S$: 
\begin{equation}
h \langle S \rangle \sim {\rm keV} \ \ \ \Longrightarrow \ \ \ 
\langle S \rangle \sim \frac{\rm keV}{h} \sim 10^2 {\rm GeV}
\end{equation}
Based on the required values of $\Omega_s$ and the mass, we conclude that
the Higgs singlet should have a VEV at the electroweak scale. In this case most
of the dark matter is produced at temperature above 100~GeV.  

At a lower temperature, some sterile neutrinos are also produced via the
Dodelson--Widrow mechanism.  This mechanism cannot be turned off. 
However, for small mixing angles, left of the yellow line in Fig.~1, the
dominant contribution comes from the $S$-boson decays.  

This has dramatic implications for the Lyman-$\alpha$ bounds because the
relation between the mass and the average momentum is very different from the
Dodelson-Widrow case.  The Lyman-$\alpha$ forest~\cite{viel} constrains the
free-streaming length of the dark-matter particles, but the relation between
this length and the particle mass depends
on the production mechanism.  One can approximately relate the free-streaming
length to the mass $m_s$ and the average momentum of the sterile neutrino: 
\begin{equation}
 \lambda_{_{FS}} \approx 1\, {\rm Mpc} \left( \frac{\rm keV}{m_s} \right)
\left(  \frac{\langle p_s \rangle}{3.15 \, T}  \right)_{T = {\rm 1\, keV}}  
\end{equation}

The sterile neutrinos produced in the $S$-boson decays have an
almost thermal spectrum at the time of production. However, as the
universe cools down, the number of effective degrees of freedom decreases from
$g_*(T_{_{S}})=110.5$ to $g_*(0.1\, {\rm MeV})=3.36$. Then $
\xi =g_*(T_{_S})/g_*(0.1\, {\rm MeV})\approx 33$. This
causes the redshifting of $\langle p_s \rangle$ by the factor $\xi^{1/3}$: 
\begin{equation}
\langle p_s \rangle_{(T \ll 1{\rm MeV})} = 0.76 \, T \left [ \frac{110.5}{
g_*(\tilde{m}_{_{S}})
}\right ]^{1/3}
\label{p_s_redshifted} 
\end{equation}
Comparing eq. (\ref{p_s_redshifted}) with the DW case, one concludes that,
as long as the population of sterile neutrinos is dominated by those produced
at a high temperature (large enough $h$, small $\theta$), the Lyman-$\alpha$
limit changes from 10~keV to 
\begin{equation}
m_s> 2.7 \, {\rm keV}
\end{equation}
This lower bound is shown in Fig.~1. 

\subsection{Sterile neutrinos and the supernova}

Sterile neutrinos with masses below several MeV can be produced in the
supernova explosion; they can play an important role in the
nucleosynthesis, as well as in generating the supernova asymmetries and the
pulsar kicks. 

The keV masses of sterile neutrinos coinside with the
Mikheev-Smirnov-Wolfenstein~\cite{msw} (MSW) resonance in nuclear matter for
typical momenta of the supernova neutrinos~\cite{ks97}.  The position of
this resonance is affected by the magnetic field due to the
D'Olivo-Nieves-Pal-Semikoz effect~\cite{ONPS}, which plays an important role
in generating the pulsar velocities, as discussed below.  Since the sterile
neutrinos interact with nuclear matter very weakly, they can be very efficient
at transporting the heat in the cooling proto-neutron star, altering the
dynamics of the supernova~\cite{Hidaka:2006sg}.  This could lead to an
enhancement of the supernova explosion. An additional enhancement can come
from the increase in convection in front of the neutron star propelled by the
asymmetric emission of sterile neutrinos~\cite{Fryer}. 

In addition to playing an important role in the primordial
nucleosynthesis~\cite{BBN,Smith:2006uw}, sterile neutrinos can affect the
$r$-process and the synthesis of heavy elements in the supernova~\cite{r}.

\subsubsection{The pulsar kicks}

The observations of
neutrinos from SN1987A constrain the amount of energy that the sterile
neutrinos can take out of the supernova, but they are still consistent with the
sterile neutrinos that carry away as much as a half of the total energy of the
supernova.  A more detailed analysis shows that the emission of sterile
neutrinos from a cooling newly born neutron star is anisotropic due to the
star's magnetic field~\cite{ks97,fkmp}.  The anisotropy of this emission can
result in a recoil velocity of the neutron star as high as $\sim 10^3$km/s.
While both the active and the sterile neutrinos are produced with some
anisotropy, the asymmetry in the amplitudes of active neutrinos is quickly
washed out in multiple scatterings as these neutrinos diffuse out of the star
in the approximate thermal equilibrium~\cite{Kusenko:1998yy}.  In contrast, the
sterile neutrinos are emitted from the supernova with the asymmetry equal to
their production asymmetry.  Hence, they give the recoiling neutron star a
momentum, large enough to explain the pulsar kicks for the
neutrino emission anisotropy as small as a few per cent~\cite{ks97,fkmp}.  
This mechanism can be the explanation of the observed pulsar 
velocities~\cite{Kusenko:review}. The range of masses and mixing angles
required to explain the pulsar kicks is shown in Fig.~\ref{fig:range}.  

The pulsar kick mechanism based on the sterile neutrino emission has several
additional predictions~\cite{Kusenko:review}: 
\begin{itemize}
 \item the kick velocities are expected to correlate with the axis of rotation 
 \item the kick should last 10 to 15 seconds, while the protoneutron star is 
cooling by the emission of neutrinos, but the onset of the kick can be delayed
by a few seconds, depending on the mass and mixing angles~\cite{fkmp}.  
\item neutrino-driven kicks can deposit additional energy behind the
supernova shock~\cite{Fryer,Hidaka:2006sg}, and they are expected to produce
asymmetric jets with the stronger jet pointing \textit{in the same direction}
as the neutron star velocity~\cite{Fryer}. 
\end{itemize}
Some of these predictions can already be tested statistically using the pulsar
data~\cite{Ng:2007aw}.  

\section{Conclusions}

The underlying physics responsible for the neutrino masses is likely to involve
the additional SU(2)-singlet fermions, or sterile neutrinos.  The Majorana
masses of these states can range from a few eV to values well above the
electroweak scale.  Theoretical arguments have been made in favor of both the
high-scale and the low-scale seesaw mechanisms: the high-scale seesaw may be
favored by the connection with the Grand Unified Theories, while the low-scale
seesaw is favored by 't~Hooft's naturalness criterion.  Cosmological
considerations are consistent with a vast range of mass scales.  The laboratory
bounds do not provide significant constraints on the sterile neutrinos, unless
they have a large mixing with the active neutrinos.  The atmospheric and solar
neutrino oscillation results cannot be reconciled with the LSND result, unless
sterile neutrinos (or other new physics) exist.  

There are several indirect astrophysical hints in favor of sterile neutrinos
at the keV scale.  Such neutrinos can explain the observed velocities of
pulsars, they can be the dark matter, and they can play a role in star
formation and reionization of the universe. 

The preponderance of indirect astrophysical hints may be a precursor of a major
discovery, although it may also be a coincidence.  One can hope to discover
the sterile neutrinos in the X-ray observations. The mass around 3~keV and the
mixing angle  $ \sin^2 \theta \sim 3\times 10^{-9}$ appear to be particularly
interesting because the sterile neutrino with such parameters could
simultaneously explain the pulsar kicks and dark matter (assuming these sterile
neutrinos are produced at the electroweak scale).  However, it is worthwhile to
search for the signal from sterile dark matter in other parts of the allowed
parameter space shown in Fig.~1.  The existence of a much lighter 
sterile neutrino, with a much greater mixing angle can be established
experimentally if MiniBooNE confirms the LSND result.

\begin{theacknowledgments}
This work was supported in part by the DOE grant DE-FG03-91ER40662 and by the
NASA ATP grants NAG~5-10842 and NAG~5-13399.  
\end{theacknowledgments}

\end{document}